\documentclass[aps,amsmath,amssymb,superscriptaddress,prx,longbibliography,twocolumn]{revtex4-2}

\usepackage{graphicx}
\usepackage{dcolumn}
\usepackage{bm}
\usepackage{xcolor}

\usepackage{ulem}
\usepackage{amsmath,amssymb}
\usepackage{multibib}
\newcites{Math}{Math Readings}

\begin{document}

\title{
Magnon-mediated qubit coupling determined via dissipation measurements
}

\author{Masaya Fukami}
\affiliation{Pritzker School of Molecular Engineering, University of Chicago, Chicago, Illinois, USA}

\author{Jonathan C. Marcks}
\affiliation{Pritzker School of Molecular Engineering, University of Chicago, Chicago, Illinois, USA}

\author{Denis R. Candido}
\affiliation{Department of Physics and Astronomy, University of Iowa, Iowa City, Iowa, USA}

\author{Leah R. Weiss}
\affiliation{Pritzker School of Molecular Engineering, University of Chicago, Chicago, Illinois, USA}
\affiliation{Advanced Institute for Materials Research, Tohoku University, Sendai 980-8577, Japan}

\author{Benjamin Soloway}
\affiliation{Pritzker School of Molecular Engineering, University of Chicago, Chicago, Illinois, USA}

\author{Sean E. Sullivan}
\affiliation{Center for Molecular Engineering and Materials Science Division, Argonne National Lab, Lemont, Illinois, USA}

\author{Nazar Delegan}
\affiliation{Center for Molecular Engineering and Materials Science Division, Argonne National Lab, Lemont, Illinois, USA}

\author{F. Joseph Heremans}
\affiliation{Pritzker School of Molecular Engineering, University of Chicago, Chicago, Illinois, USA}
\affiliation{Center for Molecular Engineering and Materials Science Division, Argonne National Lab, Lemont, Illinois, USA}

\author{Michael E. Flatt\'{e}}
\affiliation{Department of Physics and Astronomy, University of Iowa, Iowa City, Iowa, USA}
\affiliation{Department of Applied Physics, Eindhoven University of Technology, P.O. Box 513, 5600 MB Eindhoven, Netherlands}

\author{David D. Awschalom}
\email[email:]{awsch@uchicago.edu}
\affiliation{Pritzker School of Molecular Engineering, University of Chicago, Chicago, Illinois, USA}
\affiliation{Center for Molecular Engineering and Materials Science Division, Argonne National Lab, Lemont, Illinois, USA}

\begin{abstract}
Controlled interaction between localized and delocalized solid-state spin systems offers a compelling platform for on-chip quantum information processing with quantum spintronics. Hybrid quantum systems (HQSs) of localized nitrogen-vacancy (NV) centers in diamond and delocalized magnon modes in ferrimagnets--systems with naturally commensurate energies--have recently attracted significant attention, especially for interconnecting isolated spin qubits at length-scales far beyond those set by the dipolar coupling. However, despite extensive theoretical efforts, there is a lack of experimental characterization of the magnon-mediated interaction between NV centers, which is necessary to develop such hybrid quantum architectures. Here, we experimentally determine the magnon-mediated NV-NV coupling from the magnon-induced self-energy of NV centers. Our results are quantitatively consistent with a model in which the NV center is coupled to magnons by dipolar interactions. This work provides a versatile tool to characterize HQSs in the absence of strong coupling, informing future efforts to engineer entangled solid-state systems.
\end{abstract}

\maketitle

\section*{Introduction}
Quantum two-level systems coupled to bosonic fields are ubiquitous in quantum information science~\cite{main_wallraff2004strong,main_sillanpaa2007coherent,main_gustafsson2014propagating,main_tabuchi2015coherent,main_RevModPhys.59.1,main_benhelm2008towards,main_PhysRevLett.78.3086,main_rabl2010quantum}. Engineering such fundamental quantum interactions in the form of hybrid quantum systems (HQSs) provides new opportunities for quantum information processing (QIP). Among these promising hybrid quantum architectures, nitrogen-vacancy (NV) centers in diamond coupled to magnons in ferrimagnets, such as yttrium iron garnet (YIG), have recently attracted a great deal of attention~\cite{main_PhysRevX.3.041023,main_PhysRevLett.121.187204,main_candido2020predicted,main_gonzalez2022towards,main_PRXQuantum.2.040314,main_PhysRevB.105.245310,main_hetenyi2022long,main_PhysRevLett.125.247702} and have been theoretically proposed to provide opportunities for on-chip long-distance entanglement of NV centers~\cite{main_PRXQuantum.2.040314}. This is due to the NV center's desirable qubit characteristics, including long spin-coherence times~\cite{main_herbschleb2019ultra}, optical addressability~\cite{main_PhysRevLett.92.076401}, and intrinsic coupling to magnonic systems through magnetic dipolar interactions~\cite{main_PhysRevX.5.041001}. In parallel, magnonic nano- and microdevices have been extensively studied and developed in the field of magnonics for coherent transfer of spin information~\cite{main_PhysRevLett.122.247202,main_kruglyak2010magnonics,main_chumak2022advances}, paving the way for experimental realizations of proposed HQSs.

A key requirement for HQSs is strong coupling~\cite{main_wallraff2004strong,main_sillanpaa2007coherent,main_gustafsson2014propagating,main_tabuchi2015coherent} between the constituents of the system. NV-magnon coupling in general has been demonstrated experimentally using both coherent~\cite{main_andrich2017long,main_kikuchi2017long,main_bertelli2020magnetic,main_wang2020electrical,main_zhou2021magnon,main_carmiggelt2022broadband} and incoherent~\cite{main_wolfe2014off,main_van2015nanometre,main_du2017control,main_casola2018probing,main_page2019optically,main_lee2020nanoscale,main_PhysRevB.102.220403,main_purser2020spinwave,main_mccullian2020broadband,main_prananto2021probing} magnons. Surface magnons, or magnetostatic surface spin waves (MSSWs), generated coherently by microwave transducers, have been shown to efficiently drive Rabi oscillations of NV centers over long distances~\cite{main_andrich2017long,main_kikuchi2017long}. This observation implies non-zero NV-MSSW coupling~\cite{main_SI}. Additionally, thermally or incoherently driven magnons have been shown to increase the longitudinal relaxation rate and influence the optically detected magnetic resonance (ODMR) of NV centers~\cite{main_wolfe2014off,main_van2015nanometre,main_du2017control,main_casola2018probing,main_page2019optically,main_lee2020nanoscale,main_PhysRevB.102.220403,main_purser2020spinwave}, also indicative of NV-magnon coupling.

However, quantifying the magnon-mediated coupling between NV centers remains an experimental challenge. Notably, the coupling of magnons to the NV center induces a self-energy associated with the NV center \cite{main_SI}. This self-energy modifies the NV qubit's dynamics via i) a shift of the qubit energy also known as the self-energy shift and ii) a change of the qubit lifetime. Critically, the self-energy shift provides an upper-bound estimate of the magnon-mediated NV-NV coupling. Therefore, experimentally accessing the self-energy sheds light on the interactions of a given NV-magnonic HQS.

In this work, we experimentally extract the magnon-mediated coupling from the magnon-induced self-energy of NV centers interfaced with YIG. The self-energy is determined by combining room-temperature longitudinal relaxometry measurements with an analysis using the fluctuation-dissipation~\cite{main_kubo1966fluctuation} and Kramers-Kronig~\cite{main_toll1956causality} relations. Motivated by the efficient driving of NV centers via MSSWs~\cite{main_andrich2017long,main_kikuchi2017long}, we study the longitudinal relaxation of NV centers placed on top of a YIG film hosting such MSSW modes. We observe a sharply increased longitudinal relaxation rate of the NV centers driven by thermally populated MSSWs. Our experimental observations agree quantitatively with a theoretical model in which the NV center is coupled to magnons by the magnetic dipole-dipole interaction~\cite{main_PRXQuantum.2.040314}. This work builds a foundation for the hybrid quantum architecture of spin qubits coupled to magnons at the interface of quantum information science and magnonics.

\begin{figure}[t!]
\includegraphics[scale=1]{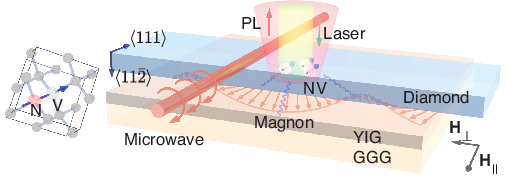}
\caption{{\bf Illustration of NV center spins interacting with magnons.} The central image shows the schematic of NV centers in a diamond slab placed on top of a YIG film. The NV axis (111) is parallel to the YIG surface, as shown in the left drawing of the diamond crystal. An in-plane external magnetic field $\mathbf{H}=\mathbf{H}_{\parallel}+\mathbf{H}_{\perp}$ is applied, where $\mathbf{H}_{\parallel}$ ($\mathbf{H}_{\perp}$) is parallel (perpendicular) to the NV axis. The perpendicular field $\mathbf{H}_{\perp}$ is zero except for Fig.~\ref{fig3}B. NV measurements are performed via confocal microscopy (laser illumination and PL collection) and pulsed microwave tones are applied with a copper wire. The thickness of the YIG film, the GGG substrate, and the diamond slab is 3~${\mathrm{\mu}}$m, 500~${\mathrm{\mu}}$m, and 100~${\mathrm{\mu}}$m, respectively. An ensemble of NV centers is implanted at approximately $7.7\pm 3.0$~nm from the bottom surface of the diamond.}
\label{fig1} 
\end{figure}

Experiments are performed on NV centers created by nitrogen implantation into a 100-${\mathrm{\mu}}$m-thick diamond slab, which is placed on top of a 3-${\mathrm{\mu}}$m-thick YIG film grown on a 500-${\mathrm{\mu}}$m-thick gadolinium gallium garnet (GGG) substrate by liquid phase epitaxy (Matesy GmbH), as illustrated in Fig.~\ref{fig1}~\cite{main_SI}. The diamond slab is laser cut (Syntek Co. Ltd) out of a bigger diamond crystal (Sumitomo) with an angle such that one of the four NV centers' main symmetry axes (NV axis) is parallel to the diamond surface as shown on the left schematic of Fig.~\ref{fig1}. The NV centers are approximately $7.7\pm 3.0$~nm from the diamond surface based on Stopping
Range of Ions in Matter (SRIM) simulations~\cite{main_ziegler2004srim}. The implanted side of the diamond slab faces down towards the YIG film in Fig.~\ref{fig1}. An ensemble of NV centers is used to increase the photoluminescence (PL) signal and consequently the signal-to-noise ratio. An external magnetic field $\mathbf{H}=\mathbf{H}_{\parallel}+\mathbf{H}_{\perp}$ is applied parallel to the diamond/YIG surface, where $\mathbf{H}_{\parallel}$ ($\mathbf{H}_{\perp}$) is parallel (perpendicular) to the NV axis. The perpendicular field $\mathbf{H}_{\perp}$ is zero except for Fig.~\ref{fig3}B. A copper wire is placed over the sample for applying microwaves to address the NV centers' electron spin transitions. A 532-nm (green) laser is focused through the diamond to the NV centers for the initialization of the NV centers' spin state to $|m_s=0\rangle$ with a confocal microscope and the PL is detected by an avalanche photodiode for the readout. Experiments are performed at room temperature and ambient conditions. 

\begin{figure}[t!]
\includegraphics{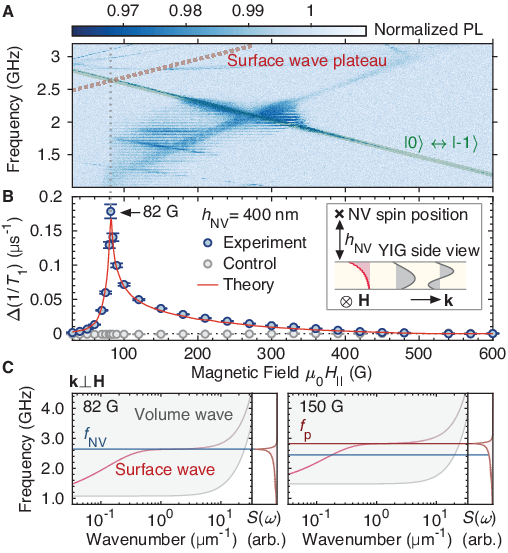}
\caption{{\bf Surface-magnon induced longitudinal relaxation.} ({\bf A}) ODMR of NV centers on YIG. The NV transition $|m_s=0\rangle\leftrightarrow |m_s=-1\rangle$ and the calculated surface spin-wave (magnon) plateau are highlighted by the solid green and the dotted red lines, respectively. The vertical dotted gray line indicates the field where the surface wave plateau is resonant with the NV transition. ({\bf B}) Longitudinal relaxation rate of NV centers for the transition $|m_s=0\rangle\leftrightarrow |m_s=-1\rangle$, which is obtained by taking the differential PL signal between measurements with and without applying a $\pi$-pulse at the end of elapsed times. The blue and the gray markers represent experiment (with YIG) and control (without YIG) measurements, respectively, and the red curve is a theoretical calculation. On the vertical axis, the $\Delta$ symbol indicates that $1/T_1$ is referenced to the value at $\mu_0 H_{\parallel}=600\ \mathrm{G}$. The NV-YIG distance is $h_{\mathrm{NV}}=400(5)\ \mathrm{nm}$ (see inset). Inset shows the cross-section of the YIG with surface (red) and volume (gray) magnon mode profiles. $\mathbf{k}$ is the wave vector. ({\bf C}) Magnon dispersion relations (for $\mathbf{k}\perp\mathbf{H}$) and corresponding noise spectra $\mathcal{S}(\omega)$ at $\mu_0 H_{\parallel}=82\ \mathrm{G}$ (left) and $\mu_0 H_{\parallel}=150\ \mathrm{G}$ (right). The surface-magnon dispersion is shown with a pink curve with the gray gradient at large $k$ indicating a suppression of the surface-wave feature, while the gray shaded area represents the band of volume waves [see inset of (B)]. The sharp peak in $\mathcal{S}(\omega)$ corresponds to the surface-wave plateau frequency $f_{\mathrm{p}}$ as indicated by the horizontal red line. The blue horizontal line denotes the NV frequency $f_{\mathrm{NV}}$.}
\label{fig2} 
\end{figure}

\section*{Results}

We show a heat map of the NV center ODMR of the system as a function of both the external magnetic field and the applied microwave frequency in Fig.~\ref{fig2}A. The distance $h_{\mathrm{NV}}=400(5)\ \mathrm{nm}$ between the NV centers and the YIG top surface is calibrated via optical interference fringes~\cite{main_SI}. While the ODMR shows multiple detailed features~\cite{main_wolfe2014off,main_lee2020nanoscale,main_SI}, including PL reduction due to transitions of off-axis NV centers and the ferromagnetic resonance of YIG, we focus on the NV centers' $|m_s=0\rangle\leftrightarrow|m_s=-1\rangle$ transition frequency $f_{\mathrm{NV}}=\omega_{\mathrm{NV}}/2\pi$ and the plateau frequency $f_{\mathrm{p}}=\omega_{\mathrm{p}}/2\pi$ of the MSSW modes, which are highlighted in Fig.~\ref{fig2}A by the solid green and the dotted red lines, respectively. Here, the magnetic field dependence of $f_{\mathrm{NV}}$ and $f_{\mathrm{p}}$ are given by $f_{\mathrm{NV}}(H)=(D_{\mathrm{NV}}-\omega_H)/2\pi$ and  $f_{\mathrm{p}}(H)=(\omega_H+\omega_M/2)/2\pi$, respectively, where $D_{\mathrm{NV}}=2\pi\times 2.87\ \mathrm{GHz}$ is the zero-field splitting of the NV center, $\omega_H=\gamma \mu_0 H$, $\omega_M=\gamma \mu_0 M_{\mathrm{s}}$, $H=|\mathbf{H}|$, $\gamma=2\pi\times 2.8\ \mathrm{MHz/G}$ is the absolute value of the electron gyromagnetic ratio, $\mu_0$ is the vacuum permeability, and $M_{\mathrm{s}}$ is the saturation magnetization of YIG. This plateau frequency $f_{\mathrm{p}}$ is the frequency of the weakly-dispersive surface-magnon mode with minimal group velocity (see Fig.~\ref{fig2}C), which coincides with the large wavenumber limit of the MSSW frequency when exchange interactions are negligible~\cite{main_SI,main_stancil2009spin}.

The longitudinal ($T_1$) relaxometry measurements are performed using the NV centers' electron spin transition $|m_s=0\rangle\leftrightarrow|m_s=-1\rangle$. To eliminate the PL contribution from off-axis NV centers, we measure the PL with and without applying a $\pi$-pulse for the $|m_s=0\rangle\rightarrow|m_s=-1\rangle$ transition at the end of a variable elapsed time $t$ between the initialization and the readout of the NV centers~\cite{main_SI}. The resulting differential PL signal is proportional to $\langle\sigma_z(t)\rangle$, where $\sigma_z=|e\rangle\langle e|-|g\rangle\langle g|$ and $|g(e)\rangle=|m_s=0(-1)\rangle$. We fit the elapsed time $t$ dependence of the differential PL signal by $\langle\sigma_z(t)\rangle/\langle\sigma_z(0)\rangle=\mathrm{exp}(-t/T_1)$ with the longitudinal relaxation time $T_1$. Magnetic field dependence of the longitudinal relaxation rate is shown in Fig.~\ref{fig2}B. Here, the rate is referenced to the value at $H_{\mathrm{r}}=600\ \mathrm{G}/\mu_0$ to eliminate the offset, i.e., $\Delta(1/T_1)\equiv 1/T_1(H)-1/T_1(H=H_{\mathrm{r}})$. This reference field is chosen because there are no resonant magnon modes for $\mu_0 H_{\mathrm{r}}\geq D_{\mathrm{NV}}/2\gamma \approx 513\ \mathrm{G}$, as the magnon-mode frequencies are lower bounded by $\omega_H$~\cite{main_SI}.

In Fig.~\ref{fig2}B, we observe a sharp peak of $\Delta(1/T_1)$ (blue markers) at the critical field $H_{\mathrm{c}}=82\ \mathrm{G}/\mu_0$ where $f_{\mathrm{NV}}(H)$ is resonant with $f_{\mathrm{p}}(H)$ (see the vertical dotted line). The inset shows a typical MSSW mode profile (red) and volume spin wave profiles (gray) to visualize the surface localization of the MSSW mode. The gray markers in Fig.~\ref{fig2}B show control measurements with NV centers in the same diamond slab without the YIG film, confirming that the peak in $\Delta(1/T_1)$ originates from the interaction between the NV centers and YIG. We compare the experimental observation with a theory in which the NV center is coupled to the dipole-exchange magnons in YIG via magnetic dipole-dipole interactions, as shown with a solid red curve in Fig.~\ref{fig2}B. Here, we apply the theory model provided in Ref.~\cite{main_PRXQuantum.2.040314} to our YIG film geometry and vary the saturation magnetization $M_{\mathrm{s}}$ as a fitting parameter, where we obtain $M_{\mathrm{s}}=1716\ \mathrm{G}/\mu_0$ (consistent with literature values)~\cite{main_SI}. We emphasize that there is no added scaling factor, indicating 
quantitative agreement between theory and experiment (see Supplementary Material~\cite{main_SI} for the goodness of the agreement).

\begin{figure}[t!]
\includegraphics[scale=1]{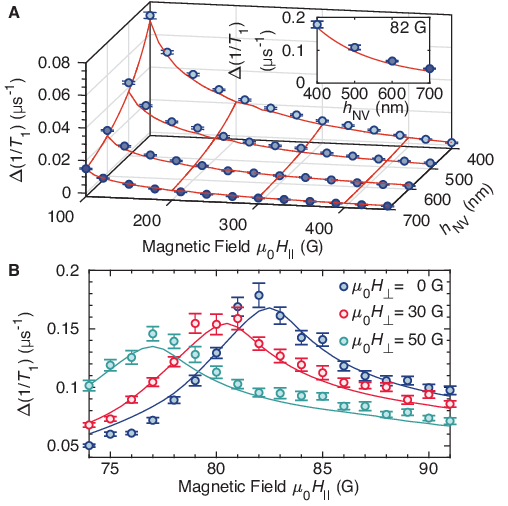}
\caption{{\bf Robust consistency between theory and experiment.} ({\bf A}) Magnetic field $H_{\parallel}$ and NV-YIG distance $h_{\mathrm{NV}}$ dependence of the longitudinal relaxation rates $\Delta(1/T_1)$. Blue markers and red curves represent experimental observations and theoretical predictions, respectively. Inset shows $h_{\mathrm{NV}}$ dependence of $\Delta(1/T_1)$ at the critical field $H_{\mathrm{c}}=82\ \mathrm{G}/\mu_0$. ({\bf B}) Parallel field $H_{\parallel}$ dependence of the longitudinal relaxation rates $\Delta(1/T_1)$ under additionally applied perpendicular fields $H_{\perp}$ (see Fig.~\ref{fig1}). Markers and curves are experimental observations and theoretical predictions, respectively, with corresponding colors.}
\label{fig3} 
\end{figure}

Dispersion relations of magnons~\cite{main_bozhko2016supercurrent,main_SI} are shown in Fig.~\ref{fig2}C under two field conditions, $82\ \mathrm{G}$ (left) and $150\ \mathrm{G}$ (right), to elaborate more on the origin of the peaked feature of $\Delta(1/T_1)$. We select the in-plane wave vector $\mathbf{k}$ of the magnon modes shown in Fig.~\ref{fig2}C to be $\mathbf{k}\perp\mathbf{H}$, the condition for MSSW modes. Dispersion relations of the surface-localized magnon modes are shown with pink curves (see the mode profiles shown in the inset of Fig.~\ref{fig2}B with corresponding colors). The corresponding magnon-induced magnetic noise spectra $\mathcal{S}(\omega)$ at the NV position are shown on the right, where $\omega$ is the angular frequency. These calculations show that the sharp peak in $\mathcal{S}(\omega)$ corresponds to $\omega=\omega_{\mathrm{p}}$ (see the horizontal red line in the right figure). As the longitudinal relaxation rate is proportional to $\mathcal{S}(\omega=\omega_{\mathrm{NV}})$, Fig.~\ref{fig2}C indicates that the peaked feature in Fig.~\ref{fig2}B is due to the combination of both the enhanced magnon-induced noise near $f_{\mathrm{p}}$ and the resonant condition between $f_{\mathrm{NV}}$ and $f_{\mathrm{p}}$. The enhanced noise spectrum near the frequency $f_{\mathrm{p}}$ is due to the maximized NV-MSSW coupling $g_{\mathbf{k},\mathrm{MSSW}}\propto \sqrt{|\mathbf{k}|}\mathrm{exp}(-|\mathbf{k}| h_{\mathrm{NV}})$ at $|\mathbf{k}|=1/2h_{\mathrm{NV}}$~\cite{main_SI} approximately corresponding to $f_{\mathrm{p}}$, as well as the large density of states of the MSSW modes near $f_{\mathrm{p}}$ (see the dispersion relation in Fig.~\ref{fig2}C).

To further substantiate the consistency between theory and experiment, we measure $\Delta(1/T_1)$ at multiple $h_{\mathrm{NV}}$ in Fig.~\ref{fig3}A for a field range $100\ \mathrm{G}\leq \mu_0 H_{\parallel}\leq450\ \mathrm{G}$, in which the NV frequency overlaps with a broad band of magnon-mode frequencies. The inset shows $h_{\mathrm{NV}}$ dependence of $\Delta(1/T_1)$ at the critical field $H_{\mathrm{c}}$. The theoretical predictions shown with the solid red curves agree well with experiments. The overall dependency on $h_{\mathrm{NV}}$ can be understood as a result of the NV-magnon coupling $g_{\mathbf{k}}\propto \mathrm{exp}(-|\mathbf{k}|h_{\mathrm{NV}})$ that decays exponentially with $h_{\mathrm{NV}}$, which originates from magnetostatics~\cite{main_SI}. We did not perform experiments with $h_{\mathrm{NV}}<400\ \mathrm{nm}$ because $\Delta(1/T_1)$ becomes larger than the initialization rate $\approx 0.2\ \mathrm{\mu s}^{-1}$ of the NV centers under the green laser illumination~\cite{main_SI}, reducing the spin polarization and the PL contrast.

Furthermore, we apply the orthogonal magnetic field $H_{\perp}$ as shown in Fig.~\ref{fig1} to investigate the response of the MSSW modes to the external magnetic field orientation~\cite{main_andrich2017long}. In Fig.~\ref{fig3}B, we show the parallel magnetic field $H_{\parallel}$ dependence (as in Fig.~\ref{fig2}B) near $H_{\mathrm{c}}$ under multiple orthogonal fields $H_{\perp}$. As the field $H_{\perp}$ is increased, we observe that $H_{\mathrm{c}}$ decreases and the amplitude of the peak height decreases. We find a good agreement between the experimental observation and the theoretical predictions as shown by the solid curves. These dependencies are qualitatively explained by the combination of the changes in the MSSW plateau frequency, the propagation direction of the MSSW, as well as the circular polarization of the magnetic noise~\cite{main_PhysRevB.102.220403,main_casola2018probing} generated by the MSSWs~\cite{main_SI}.

\begin{figure}[t!]
\includegraphics[scale=1]{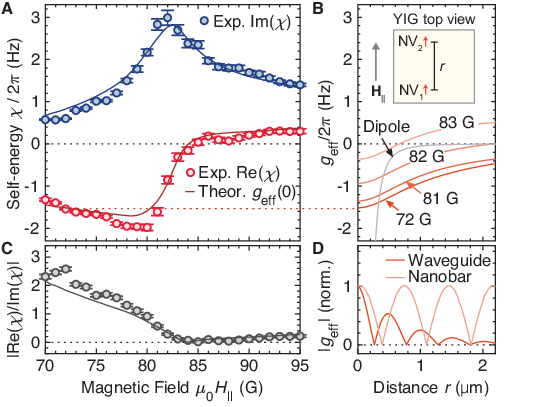}
\caption{{\bf Magnon-mediated coupling determined from magnon-induced self-energy.} ({\bf A}) Magnon-induced NV center self-energy $\chi$ as a function of the magnetic field $H_{\parallel}$. The red (blue) markers represent the experimentally determined real (imaginary) part of $\chi$. The red and blue curves show the theoretical predictions of $g_{\mathrm{eff}}(r=0)=\mathrm{Re}(\chi)$ and $\mathrm{Im}(\chi)$, respectively. ({\bf B}) Calculated NV-NV distance $r$ dependence of the effective NV-NV coupling strength $g_{\mathrm{eff}}(r)$ mediated by magnons at multiple fields $H_{\parallel}$. Two NV centers are placed at the same distance $h_{\mathrm{NV}}=400\ \mathrm{nm}$ from the YIG surface and displaced in the direction of $\mathbf{H}_{\parallel}$ as shown in the top illustration. The dipole-dipole interaction strength $g_{\mathrm{dip}}(r)$ is shown with a gray curve as a reference. ({\bf C}) Ratio of the real and imaginary parts of $\chi$ as the magnetic field is swept through the MSSW resonance, indicating how the ratio of NV-NV coupling to dissipation varies with the NV frequency. Markers and the gray curve are experimental and theoretical results, respectively. ({\bf D}) Calculated $g_{\mathrm{eff}}$ as a function of $r$ in different geometries of YIG, an infinitely-long waveguide and a nanobar. Two curves are calculated for the identical conditions as in Ref.~\cite{main_PRXQuantum.2.040314}. Vertical axis is normalized by $g_{\mathrm{eff}}(r=0)$.}
\label{fig4} 
\end{figure}

To characterize the NV-magnon HQS, we determine both the real ($\chi^{\prime}$) and imaginary ($\chi^{\prime\prime}$) parts of the self-energy $\chi=\chi^{\prime}+i\chi^{\prime\prime}$ originating from the NV centers' interaction with magnons. It modifies the NV frequency as $\omega_{\mathrm{NV}}\rightarrow \omega_{\mathrm{NV}}-\chi$, indicating that the time evolution of the NV center wave function is modified to $\psi(t>0)\propto \mathrm{exp}[-i(\omega_{\mathrm{NV}}-\chi)t]=\mathrm{exp}(-\chi^{\prime\prime}t)\mathrm{exp}[-i(\omega_{\mathrm{NV}}-\chi^{\prime})t]$~\cite{main_SI}. Hence, $\chi^{\prime}$ and $\chi^{\prime\prime}$ represent the self-energy shift [or Lamb shift~\cite{main_breuer2002theory}] and the decay rate of the NV center spin, respectively. While the strength of $\chi$ in our system is small due to the lack of magnon confinement in the YIG film, making it hard to measure directly, we are able to determine it by taking advantage of the fundamental fluctuation-dissipation and Kramers-Kronig relations~\cite{main_kubo1966fluctuation,main_toll1956causality}. To this end, for the imaginary part we use the fluctuation-dissipation relation~\cite{main_kubo1966fluctuation,main_SI}
\begin{eqnarray}
\chi^{\prime\prime}(H)=\frac{\Delta(1/T_1)}{2\mathrm{coth}(\beta\omega_{\mathrm{NV}}/2)},\label{ImPartFDR}
\end{eqnarray}
where $\beta=1/k_{\mathrm{B}}T$ is the inverse temperature, $k_{\mathrm{B}}$ is the Boltzman constant, $T=299.6(3)\ \mathrm{K}$ is the temperature measured by the thermometer attached to the sample base, and we set $\hbar=1$. At room temperature, the denominator on the right-hand side of Eq.~(\ref{ImPartFDR}) becomes $2\mathrm{coth}(\beta\omega_{\mathrm{NV}}/2)\approx 4k_{\mathrm{B}} T/\omega_{\mathrm{NV}}$, indicating that the effect of $\chi^{\prime\prime}$ on $1/T_1$ is amplified by the temperature. This enables us to experimentally probe the small magnon-induced self-energy at room temperature. For the real part, we use the Kramers-Kronig relation~\cite{main_toll1956causality,main_SI}
\begin{eqnarray}
\chi^{\prime}(H)=\mathcal{P}\int \frac{dH^{\prime}}{\pi}\frac{\chi^{\prime\prime}(H^{\prime})}{H-H^{\prime}},\label{RePartKKR}
\end{eqnarray}
where $\mathcal{P}$ indicates the Cauchy principal value. Applying Eqs.~(\ref{ImPartFDR}) and (\ref{RePartKKR}) to the measurements of $\Delta(1/T_1)$, we obtain both the real and imaginary parts of $\chi(H)$.

Fig.~\ref{fig4}A shows the real and imaginary parts of $\chi(H)$ as a function of the external field. Importantly, the real part $\chi^{\prime}$ provides an upper-bound estimate of the magnon mediated NV-NV interaction $g_{\mathrm{eff}}$, which can be tuned by changing the external magnetic field by a few Gauss~\cite{main_PRXQuantum.2.040314}. More accurately, it provides $g_{\mathrm{eff}}(r=0)$, where $r$ is the distance between two NV centers placed on top of the YIG film (see the illustration in Fig.~\ref{fig4}B). This is because $\chi^{\prime}$ can be regarded as a special case of the two-qubit interaction (between qubits $i$ and $j$) where the two qubits are in fact identical ($i=j$)~\cite{main_SI}. The theoretical prediction of $g_{\mathrm{eff}}(r=0)$ is shown in Fig.~\ref{fig4}A with a solid red curve to be contrasted with $\chi^{\prime}$ (also see Fig.~\ref{fig4}B at $r=0$ and the horizontal dotted line). For magnon-mediated two-qubit gates, it is desirable to work with small $\chi^{\prime\prime}$ and large $|\chi^{\prime}|$ by adding a detuning from the resonance condition~\cite{main_PRXQuantum.2.040314}, such as $\mu_0 H_{\parallel}\leq80\ \mathrm{G}$. In Fig.~\ref{fig4}C we show $|\chi^{\prime}/ \chi^{\prime \prime}|$ as a function of the external magnetic field, which quantifies the relation between the magnon-mediated two-qubit coupling $g_{\mathrm{eff}}(0)=\chi^{\prime}$ and the magnon-limited decoherence rate at low temperature $\chi^{\prime\prime}$. The figure shows that the increase in dissipation as the magnetic field is tuned closer to resonance is a much larger effect than the modest increase in coupling strength. We observed as large as $|\chi^{\prime}/ \chi^{\prime \prime}|\approx 2.5$ near $72\ \mathrm{G}$, showing that the magnon-mediated coupling in principle allows for a useful two-qubit gate by overcoming the decoherence rate if the decoherence is limited only by magnons. Although this is not the case in our system as the NV center's coherence time~\cite{main_SI} is much shorter than $1/\chi^{\prime\prime}$, we expect it to be achievable in systems with a lower nitrogen concentration~\cite{main_PhysRevB.102.134210} and larger NV-magnon coupling at low temperature. We further note that the time required for the $\sqrt{i\rm SWAP}$ entangling gate is $\tau_{\sqrt{i\mathrm{SWAP}}}=(1/8)\times 2\pi/|g_{\mathrm{eff}}|$~\cite{main_PRXQuantum.2.040314}. This implies that the magnon-limited gate-to-decoherence ratio $(\mathrm{GDR})$~\cite{main_PRXQuantum.2.040314} in our system near $72\ \mathrm{G}$ is $(\mathrm{GDR})=(4/\pi) \times |\chi^{\prime}/ \chi^{\prime \prime}|\approx 3$. 

We show in Fig.~\ref{fig4}B the calculated $r$ dependence of $g_{\mathrm{eff}}(r)$ at multiple magnetic fields, where $g_{\mathrm{eff}}(0)$ varies significantly, to verify that $\chi^{\prime}=g_{\mathrm{eff}}(0)$ provides an upper-bound estimate of $g_{\mathrm{eff}}(r)$. It shows that the magnitude of $g_{\mathrm{eff}}(r)$ is typically the largest at $r=0$, indicating $|g_{\mathrm{eff}}(r)|\lesssim|g_{\mathrm{eff}}(0)|=|\chi^{\prime}|$. We note that the distance $r$ dependence of $g_{\mathrm{eff}}$ is highly geometry dependent as shown in Fig.~\ref{fig4}D. In the figure, we calculate $g_{\mathrm{eff}}$ in two different YIG structures, an infinite waveguide and a nanobar, for the identical conditions as calculated in Ref.~\cite{main_PRXQuantum.2.040314}. While the coupling in the infinite waveguide drops off at length-scales relevant for long-distance entanglement, the nanobar sustains the coupling strength over two micrometers. We experimentally obtain $|\chi^{\prime}|\approx2\pi\times 2\ \mathrm{Hz}$ in Fig.~\ref{fig4}A, which is notably greater than the bare magnetic dipole-dipole interaction $|g_{\mathrm{dip}}|$ at $r=0.5\ {\mathrm{\mu}}\mathrm{m}$ [see the gray curve in Fig.~\ref{fig4}B with $|g_{\mathrm{dip}}(r=0.5\ {\mathrm{\mu}}\mathrm{m})|\approx 2\pi\times 0.4\ \mathrm{Hz}$]. The condition $|g_{\mathrm{eff}}(r)|\gg |g_{\mathrm{dip}}(r)|$ is necessary for the magnon-mediated NV-NV interaction to be useful in QIP~\cite{main_SI}. While these results are promising, more generally, this analysis procedure is widely applicable to any hybrid quantum architecture with qubits coupled to magnons, while being particularly validated for our system where the MSSW modes play a major role~\cite{main_SI}.

\section*{Discussion}
To quantify the NV-magnon interaction of the system, we define a dimensionless parameter $\mathcal{C}\equiv \chi^{\prime\prime}T_2^*$, where $T_2^*$ is the Ramsey decoherence time of the NV center~\cite{main_SI}. This parameter becomes the cooperativity~\cite{main_PhysRevLett.118.223603,main_PhysRevLett.115.063601,main_chang2007single} if we have a single magnon mode, i.e. our definition is a generalization of the cooperativity. In our experiments, decoherence is dominantly caused by nearby P1 centers~\cite{main_PhysRevB.102.134210} which leads to a short Ramsey decay time $T_2^{*}\approx 180\ \mathrm{ns}$, resulting in $\mathcal{C}\approx 3\times10^{-6}$. However, assuming a single NV center with a long coherence time $T_{2,\mathrm{ref}}^*\approx1\ \mathrm{ms}$~\cite{main_herbschleb2019ultra}, the projected value of the parameter is $\mathcal{C}_{\mathrm{proj}}\approx 2\times 10^{-2}$~\cite{main_SI}. Furthermore, based on the scaling~\cite{main_candido2021theory} of $\Delta(1/ T_1)\propto h_{\mathrm{NV}}^\alpha$ with $\alpha=-2.4(1)$ obtained by fitting the inset of Fig.~\ref{fig3}B~\cite{main_SI}, an approximately 250-fold enhancement in $\mathcal{C}$ and in $g_{\mathrm{eff}}$ are estimated when $h_{\mathrm{NV}}$ is smaller by a factor of ten. To explore this smaller $h_{\mathrm{NV}}$ regime, however, room-temperature experiments result in too large longitudinal relaxation rates, e.g., $1/T_1\approx 250\times 0.18\ \mathrm{\mu s}^{-1}\approx 45\ \mathrm{\mu s}^{-1}$ ($T_1\approx 22\ \mathrm{ns}$) at $82\ \mathrm{G}$. Therefore, experiments need to be performed at lower temperatures to suppress the thermal magnon occupation $n_\mathrm{B}=1/(\mathrm{exp(\beta \omega_{\mathrm{NV}})}-1)$ [see Eq.~(\ref{ImPartFDR}) indicating $\Delta(1/T_1)=2\chi^{\prime\prime}\times (2n_\mathrm{B} +1)$ and recall the limitation of the $T_1$ measurement due to the initialization rate]. For example, at $T=4\ \mathrm{K}$, it is anticipated that $T_1$ is extended approximately by a factor of $75$ ($T_1\approx 1.7\ \mathrm{\mu s}$ at $82\ \mathrm{G}$) as compared to the room-temperature value, making it more experimentally accessible to polarize the NV centers optically. Importantly, even larger cooperativities are expected when magnonic micro- or nanostructures are used~\cite{main_chumak2022advances}, motivating further experimental studies of the HQSs of NV centers and magnons. 

In conclusion, we experimentally quantify the magnon-induced self-energy of NV centers to characterize the interaction strength between NV centers and magnons in a HQS. We observe an increased longitudinal relaxation rate of the NV centers caused by thermal surface magnons, which is complementary to Refs.~\cite{main_andrich2017long,main_kikuchi2017long}. With this observation, the upper-bound of the surface-magnon mediated NV-NV interaction in our system is estimated to be $|g_{\mathrm{eff}}|\approx2\pi \times 2\ \mathrm{Hz}$. The experimental observation agrees quantitatively with a theory model developed in Ref.~\cite{main_PRXQuantum.2.040314} after the current sample geometry is taken into account. By combining room-temperature longitudinal relaxometry measurements with an analysis using the fluctuation-dissipation and Kramers-Kronig relations, our characterization approach does not require millikelvin temperatures needed for entanglement generation~\cite{main_PRXQuantum.2.040314}, simplifying experimental implementation. This work provides a versatile method to leverage incoherent interactions in weakly coupled systems to characterize HQS platforms.

\section*{Acknowledgments}
We thank Xinghan Guo for preliminary sample preparations and Paul C. Jerger for helpful discussions. This work was primarily supported by the U.S. Department of Energy, Office of Science, Basic Energy Sciences, Materials Sciences and Engineering Division (S.E.S., N.D., F.J.H., D.D.A.) with additional support from Q-NEXT, a U.S. Department of Energy Office of Science National Quantum Information Science Research Centers (M.F. J.C.M.), and the Air Force Office of Scientific Research (M.F., D.D.A.). This work made use of Pritzker Nanofabrication Facility, which receives support from the SHyNE; a node of the NSF’s National Nanotechnology Coordinated Infrastructure (NSF ECCS-1542205). J.C.M. acknowledges prior support from the National Science Foundation Graduate Research Fellowship Program (grant no. DGE-1746045). L.R.W. acknowledges support from the University of Chicago/Advanced Institute for Materials Research Joint Research Center. M. E. F. and D. R. C. acknowledge support by the U.S. Department of Energy, Office of Science, Basic Energy Sciences under Award No. DE-SC0019250.

\normalem

\end{document}